\documentclass[prb,twocolumn,superscriptaddress,showpacs, nofootinbib]{revtex4}
\usepackage{graphicx}
\usepackage{float}
\usepackage{dcolumn}
\usepackage{float}
\usepackage{bm}
\usepackage{mathrsfs}
\usepackage{ulem}
\usepackage[per-mode = fraction]{siunitx}
\usepackage{hyperref}
\usepackage{xcolor}
\usepackage{soul}
\usepackage{amssymb}
\usepackage{amsthm}
\usepackage{mathtools}
\begin{document}

\title{Magneto-topological transitions in multicomponent superconductors}

\author{Yuriy Yerin}
\affiliation{Dipartimento di Fisica e Geologia, Universit\'a degli Studi di Perugia, Via Pascoli, 06123 Perugia, Italy}
\author{Stefan-Ludwig Drechsler}
\affiliation{Institute for Theoretical Solid State Physics, Leibniz-Institut für Festkörper- und Werkstoffforschung IFW-Dresden, D-01169 Dresden, Helmholtzstraße 20}
\author{Mario Cuoco}
\affiliation{CNR-SPIN, c/o Universit\'a di Salerno, I-84084 Fisciano (SA), Italy}
\author{Caterina Petrillo}
\affiliation{Dipartimento di Fisica e Geologia, Universit\'a degli Studi di Perugia, Via Pascoli, 06123 Perugia, Italy}
\date{\today}

\begin{abstract}
Multi-component spin-singlet superconductors with competing 0- and $\pi$-pairing couplings, as in $s_{++}$ and $s_{\pm}$ phases, are close to instabilities with a spontaneous breaking of time-reversal symmetry.  
We demonstrate that the modification of the kinetic energy of superconducting electrons in a doubly connected superconducting cylinder, determined by the applied flux, generally drives transitions from chiral superconducting states to configurations that are time-reversal symmetric. This magneto-topological-induced changeover is investigated by means of a Ginzburg-Landau approach for a two-band superconductor with interband interactions and impurity scattering investigated for the case of a sample in the form of a mesoscopically thin-walled cylinder. We find that the application of a magnetic flux can convert a chiral $s_{\pm}+is_{++}$ state into a $s_{\pm}$ configuration and vice versa or tune the energy splitting of chiral states having inequivalent pairing amplitudes. We discuss signatures for the detection of these phases and of the corresponding transitions in mesoscopic superconducting loops.
\end{abstract}
\maketitle

\section{Introduction}
One of the major challenges in condensed matter physics is to unravel the fundamental structure of the electron pairing in unconventional superconductors. This problem is of special relevance for correlated electrons materials where pairing with either breaking of time reversal or inversion symmetry can occur. Paradigmatic examples in this context are represented by strontium ruthenate \cite{Maeno, Kallin}, iron-based \cite{Tafti, Mazin, Klauss}, noncentrosymmetric \cite{Smidman} and heavy-fermions superconductors \cite{Izawa}. 

Since most unconventional superconductors are marked by a multi-orbital electronic structure, emergent anomalous behaviors are expected as due to the multi-component character of the superconducting order parameter. 
A typical manifestation is given by intrinsic $\pi$-phase shift or $\pi$-pairing, i.e. an anti-phase relation between the superconducting order parameters in different bands.
This type of band-dependent phase rearrangement is at the heart of unconventional superconductivity in iron-based \cite{Klauss,Gri21}, oxide interface superconductors \cite{Sche15,Sin22}, electrically or orbitally driven superconducting phases \cite{Mer20,Bou20,DeSim21,Mer21}, and multi-orbital noncentrosymmetric superconductors \cite{Fuk18,Fuk20,Sche15,Mer20}.

Clear-cut challenges in this framework are to assess whether the superconducting phase frustration in the presence of competing $0$ and $\pi$-pairings leads to time-reversal symmetry breaking~\cite{Gri20,Gri21,Tri21} and, in turn, to single out specific detection schemes for accessing the complexity of multi-component superconductors. 

To these aims, in this Letter we demonstrate that for a superconductor with competing pairing channels with $0$- and $\pi$-coupling, the response to an external magnetic flux, in a suitably designed non-simple connected mesoscopic geometry (see Fig. \ref{cylinder_thin}), generally leads to transitions from phases with broken time-reversal symmetry (BTRS) to time-reversal symmetry conserving states. The analysis is based on a two-band superconducting model whose repulsive interband interactions and interband impurity scattering set out a chiral phase with the chiral order parameter having $s_{\pm}+is_{++}$  symmetry. We unveil how the modification of the kinetic energy of the superconducting electrons in a doubly connected superconducting cylinder drives a transition between chiral phases and time-reversal conserving configurations with $\pi$-pairing ($s_{\pm}$). Interestingly, the application of the magnetic flux can also tune the energy difference between chiral phases with a different amplitude of the superconducting order parameter. These findings are characteristic of any configuration with non-simple connected geometry and indicate a general transition behavior when a superconductor, with time-reversal symmetry breaking associated to a phase frustration of the internal degrees of freedom, is subjected to a magnetic flux in a superconducting ring.  

\section{Formalism and methodology}
We use the Ginzburg-Landau (GL) theory applied to a dirty two-band superconductor. For this physical case, by means of the Usadel equations one can deduce the Gibbs free energy $G$ \cite{Stanev, Corticelli} which is generally expressed as
\begin{equation}
\label{GL_general}
G = {F_1} + {F_2} + {F_{12}} + \int{\frac{{{{\left( {{\text{rot }}{\mathbf{A}} - {\mathbf{H}}} \right)}^2}}}{{8\pi }}} {d^3}{\mathbf{r}},
\end{equation}
where $F_i$ are the partial contributions of the \textit{i}th band, $F_{12}$ is the component arising from the interband interaction which is also affected by the presence of interband impurity scattering. The last term describes the contribution of an external magnetic field. The expressions for $F_i$ and $F_{12}$ are provided below
\begin{widetext}
\begin{equation}
\label{GL_general1}
{F_1} = \int{\left[ {{a_{11}}{{\left| {{\Delta _1}} \right|}^2} + \frac{1}{2}{b_{11}}{{\left| {{\Delta _1}} \right|}^4} + \frac{1}{2}{k_{11}}{{\left| { - i\hbar \nabla  - \frac{{2e}}{c}{\mathbf{A}}} \right|}^2}{\Delta _1}} \right]} {d^{\mathbf{3}}}{\mathbf{r}},
\end{equation}
\begin{equation}
\label{GL_general2}
{F_2} = \int{\left[ {{a_{22}}{{\left| {{\Delta _2}} \right|}^2} + \frac{1}{2}{b_{22}}{{\left| {{\Delta _2}} \right|}^4} + \frac{1}{2}{k_{22}}{{\left| { - i\hbar \nabla  - \frac{{2e}}{c}{\mathbf{A}}} \right|}^2}{\Delta _2}} \right]} {d^{\mathbf{3}}}{\mathbf{r}},
\end{equation}
\begin{equation}
\label{GL_general3}
\begin{gathered}
  {F_{12}} = \int{\left[ {{b_{12}}{{\left| {{\Delta _1}} \right|}^2}{{\left| {{\Delta _2}} \right|}^2}} \right.}  + 2\left( {{a_{12}}\left| {{\Delta _1}} \right|\left| {{\Delta _2}} \right| + {c_{11}}{{\left| {{\Delta _1}} \right|}^3}\left| {{\Delta _2}} \right| + {c_{22}}\left| {{\Delta _1}} \right|{{\left| {{\Delta _2}} \right|}^3}} \right)\cos \phi  + {c_{12}}{\left| {{\Delta _1}} \right|^2}{\left| {{\Delta _2}} \right|^2}\cos 2\phi  \hfill \\
  \left. { + \frac{1}{2}{k_{12}}\left( {\left( { - i\hbar \nabla  - \frac{{2e}}{c}{\mathbf{A}}} \right){\Delta _1}\left( {i\hbar \nabla  - \frac{{2e}}{c}{\mathbf{A}}} \right)\Delta _2^* + \left( {i\hbar \nabla  - \frac{{2e}}{c}{\mathbf{A}}} \right)\Delta _1^*\left( { - i\hbar \nabla  - \frac{{2e}}{c}{\mathbf{A}}} \right){\Delta _2}} \right)} \right]{d^3}{\mathbf{r}}. \hfill \\ 
\end{gathered} 
\end{equation}
\end{widetext}
Here, ${\Delta _i} = \left| {{\Delta _i}} \right|\exp \left( {i{\chi _i}} \right)$ are complex order parameters and $\phi  = {\chi _2} - {\chi _1}$ is the phase difference. The coefficients of the Gibbs free energy functional are reported in the in Appendix \ref{sec:A}. The coefficients $b_{12}$, $c_{ij}$ and $k_{12}$ in Eq. (\ref{GL_general3}) are absent in the case of a clean two-band superconductor. They are a direct consequence of the contribution of the interband impurities, whose strength is characterized by the interband scattering rate $\Gamma$, being proportional to the impurity concentration.

\begin{figure}
\includegraphics[width=0.99\columnwidth]{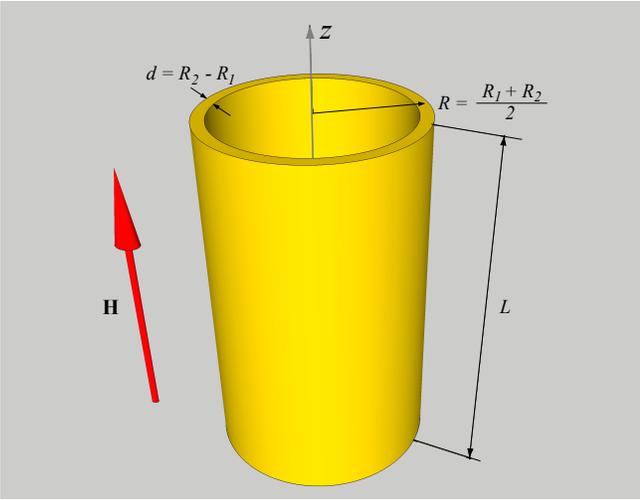}
\caption {Sketch of the geometrical configuration for the examined problem \onlinecite{Yerin1, Yerin2} with a thin cylinder. $H$ is the applied magnetic field along the $z$-axis of the cylinder. The ring has an internal (external) radius which is given by $R_1$ ($R_2$), respectively.}
\label{cylinder_thin}
\end{figure}

The main idea behind the magneto-topological transitions is to exploit a combined use of doubly connected topology and external magnetic field. To this end, as an illustrative example of such physical scenario we consider a long tube ($L$ is the length) with a thin wall, with a thickness $d$ that is assumed to be much smaller than the characteristic coherence length(es) $\xi_1$, $\xi_2$, while the radius $R = \frac{{{R_1} + {R_2}}}{2}$ has to be larger (Figure \ref{cylinder_thin}). When the condition $\frac{{dR}}{{2{\lambda ^2}}} \ll 1$  is fulfilled, where $\lambda$ is the weak-field penetration depth, the Meissner effect is small (for more details see Ref. \onlinecite{Yerin6}). The cylindrical coordinates  ($r,\varphi ,z$) are introduced, where the \textit{z} axis coincides with the axis of a cylinder.  The constant external magnetic field $H$ is applied along the symmetry axis with the vector potential ${\mathbf{A}} = \left( {0,{A_\varphi }\left( r \right),0} \right),{A_\varphi }\left( r \right) = \frac{{Hr}}{2}$ (Fig. (\ref{cylinder_thin})). This allows us to neglect the\textit{r}-and \textit{z} dependencies of the order parameter, which are relevant for thick short tubes. Also, these conditions preclude the formation of vortices in the wall of the cylinder and guarantee that self-induced magnetic fields are small. 

Bearing in mind the doubly connected topology of the superconductor, we diagonalize the Gibbs free energy and reduce it to the following expression (see details of the derivation in Appendix \ref{sec:B})
\begin{widetext}
\begin{equation}
\label{Gibbs_final}
\begin{gathered}
  \frac{G}{{{V_s}}} = {F_0} + \left[ {\left( {\frac{1}{2}{k_{11}}{{\left| {{\Delta _1}} \right|}^2} + \frac{1}{2}{k_{22}}{{\left| {{\Delta _2}} \right|}^2} + {k_{12}}\left| {{\Delta _1}} \right|\left| {{\Delta _2}} \right|\cos \phi } \right){{\hbar ^2}}}{{}}{{q}^2} \right. \hfill \\
  \left. { + 2\left( {{a_{12}}\left| {{\Delta _1}} \right|\left| {{\Delta _2}} \right| + {c_{11}}{{\left| {{\Delta _1}} \right|}^3}\left| {{\Delta _2}} \right| + {c_{22}}\left| {{\Delta _1}} \right|{{\left| {{\Delta _2}} \right|}^3}} \right)\cos \phi  + {c_{12}}{{\left| {{\Delta _1}} \right|}^2}{{\left| {{\Delta _2}} \right|}^2}\cos 2\phi } \right], \hfill \\ 
\end{gathered}
\end{equation}
\end{widetext}
where ${V_s} = 2\pi RLd$ is the volume of the material part of a cylinder and $F_0$ is
\begin{equation}
\label{F0}
\begin{gathered}
  {F_0} = {a_{11}}{\left| {{\Delta _1}} \right|^2} + {a_{22}}{\left| {{\Delta _2}} \right|^2} + \frac{1}{2}{b_{11}}{\left| {{\Delta _1}} \right|^4} \\ 
   + \frac{1}{2}{b_{22}}{\left| {{\Delta _2}} \right|^4} + {b_{12}}{\left| {{\Delta _1}} \right|^2}{\left| {{\Delta _2}} \right|^2}. \\ 
\end{gathered} 
\end{equation}
Here, we introduce the wave-vector $q (\Phi) = \frac{1}{R}\mathop {\min }\limits_N \left({N - \frac{\Phi }{{{\Phi _0}}}} \right)$, which is expressed through the winding number $N$.  The winding number $N$ arises from the topological properties of the cylinder (its double-connectedness) and the quantization rule for the order parameter phases
\begin{equation}
\label{quantization_rule}
\oint\limits_C {\nabla {\chi _i} \cdot d{\mathbf{l}}}  = 2\pi {N_i},
\end{equation}
where $C$ is an arbitrarily closed contour that lies inside the wall of the cylinder and encircles the opening and $N_i =0, \pm 1, \pm 2,...$ are winding numbers for $i$-th component of the order parameter. The expression for the Gibbs free energy Eq. (\ref{Gibbs_final}) is obtained within the assumption of a homogeneous state, i.e., $N_1=N_2=N$, taking into account the symmetry of the problem and the continuity conditions. We will not consider different inhomogeneous solutions for the examined problem when $N_1 \ne N_2$ (see Appendix \ref{sec:B}). We  note that inhomogeneities add extra complexity to the problem as several unconventional states can arise. 
For instance, in the bulk of a multi-component superconductor fractional vortices can occur \cite{Corticelli, Babaev2002, Silaev2013, Tanaka1, Tanaka2, Tanaka3}, while in the case of a doubly-connected topology, with magnetic vortices in the volume of the superconductor being energetically unfavorable, inhomogeneous state of solitons type can form \cite{Tanaka2002, Bluhm, Yerin6, Vakaryuk}. Solitons also occur in the case of planar geometry generating a phase kink of the sine-Gordon type \cite{Lin2, Samokhin2, Arisawa, Vodolazov}. Moreover, some inhomogeneous solutions are marked by non-equilibrium phase textures \cite{Gurevich2003, Yerin5}, domain walls \cite{Babaev5}, or unusual Fulde-Ferrell-Larkin-Ovchinnikov (FFLO) pairing \cite{Ptok, Machida1, Machida2} and other configurations arise from the interplay of the geometry of the superconductor and the spatial dependence of the magnetic field in the superconductor \cite{Hayashi, Askerzade1, Yerin3}.

The calculation of the functional derivatives $\partial G/\partial \phi = 0$, $\partial G/\partial |\Delta_1| = 0$ and $\partial G/\partial |\Delta_2| = 0$ leads to equations for $|\Delta_i|$ and allows us to obtain solutions for the parameter $\phi$ (see details of the derivation in Appendix \ref{sec:B})
\begin{equation}
\label{phi_homo_sol1}
\sin \phi  = 0 \Rightarrow \phi  = 0,\phi  = \pi,
\end{equation}
which corresponds to $s_{++}$ and $s_{\pm}$ symmetry, respectively. The most interesting case is the BTRS solution with an arbitrary $\phi$ and the accompanied chiral symmetry $s_{\pm}+is_{++}$ 
\begin{equation}
\label{phi_homo_sol2}
\cos \phi  =  - \frac{{{{k_{12}}{\hbar ^2}}}{{}{{q}^2} + 2\left( {{a_{12}} + {c_{11}}{{\left| {{\Delta _1}} \right|}^2} + {c_{22}}{{\left| {{\Delta _2}} \right|}^2}} \right)}}{{4{c_{12}}\left| {{\Delta _1}} \right|\left| {{\Delta _2}} \right|}},
\end{equation}
which gives rise to two solutions for the phase difference and consequently leads to a sort of frustration with two degenerate ground states and spontaneously broken ${\mathbb{Z}_2}$ time-reversal symmetry.

For $q=0$ and for the BTRS states, one can derive analytical solutions for the amplitudes of the superconducting order parameters. There are two solutions which are expressed as
\begin{widetext}
\begin{equation}
\label{OP1_sol1}
\left| {\Delta _1^{\left( 0 \right)}} \right| = \sqrt { - \frac{{{a_{11}}{b_{22}}{c_{12}} - {a_{11}}c_{22}^2 + {a_{12}}{b_{12}}{c_{22}} - {a_{12}}{b_{22}}{c_{11}} - {a_{12}}{c_{12}}{c_{22}} - {a_{22}}{b_{12}}{c_{12}} + {a_{22}}{c_{11}}{c_{22}} + {a_{22}}c_{12}^2}}{{{b_{11}}{b_{22}}{c_{12}} - {b_{11}}c_{22}^2 - b_{12}^2{c_{12}} + 2{b_{12}}{c_{11}}{c_{22}} + 2{b_{12}}c_{12}^2 - {b_{22}}c_{11}^2 - 2{c_{11}}{c_{12}}{c_{22}} - c_{12}^3}}},
\end{equation}
\begin{equation}
\label{OP2_sol1}
\left| {\Delta _2^{\left( 0 \right)}} \right| = \sqrt {\frac{{{a_{11}}{b_{12}}{c_{12}} - {a_{11}}{c_{11}}{c_{22}} - {a_{11}}c_{12}^2 + {a_{12}}{b_{11}}{c_{22}} - {a_{12}}{b_{12}}{c_{11}} + {a_{12}}{c_{11}}{c_{12}} - {a_{22}}{b_{11}}{c_{12}} + {a_{22}}c_{11}^2}}{{{b_{11}}{b_{22}}{c_{12}} - {b_{11}}c_{22}^2 - b_{12}^2{c_{12}} + 2{b_{12}}{c_{11}}{c_{22}} + 2{b_{12}}c_{12}^2 - {b_{22}}c_{11}^2 - 2{c_{11}}{c_{12}}{c_{22}} - c_{12}^3}}},
\end{equation}
\end{widetext}
and
\begin{equation}
\label{OP1_sol2}
\left| {\Delta _1^{\left( 0 \right)}} \right| = \sqrt { - \frac{{{a_{11}}{c_{12}} - {a_{12}}{c_{11}}}}{{{b_{11}}{c_{12}} - c_{11}^2}}},
\end{equation}
\begin{equation}
\label{OP2_sol2}
\left| {\Delta _2^{\left( 0 \right)}} \right| = \sqrt {\frac{{{a_{12}}{c_{22}} - {a_{22}}{c_{12}}}}{{{b_{22}}{c_{12}} - c_{22}^2}}}.
\end{equation}

The subsequent substitution of the expression for the phase difference in the BTRS state given by Eq. (\ref{phi_homo_sol2}) into Eq. (\ref{Gibbs_final}) yields the following fourth-order polynomial of $q$
\begin{widetext}
\begin{equation}
\label{Gibbs_final_new}
\begin{gathered}
  \frac{G}{{{V_s}}} = {F_0} - \frac{1}{{{c_{12}}}}\left[ {\frac{{{k_{12}^2\hbar ^4{q}^4}}}{{8{}}}{}} \right.
   + \left( {\left( {{c_{11}}{k_{12}} - {c_{12}}{k_{11}}} \right){{\left| {{\Delta _1}} \right|}^2} + \left( {{c_{22}}{k_{12}} - {c_{12}}{k_{22}}} \right){{\left| {{\Delta _2}} \right|}^2} + {a_{12}}{k_{12}}} \right)\frac{{{\hbar ^2q^2}}}{{2{}}}{} \hfill \\
  \left. { + \left( {\frac{1}{2}{a_{12}} + {c_{11}}{{\left| {{\Delta _1}} \right|}^2} + {c_{22}}{{\left| {{\Delta _2}} \right|}^2}} \right){a_{12}} + \frac{1}{2}{{\left( {{c_{11}}{{\left| {{\Delta _1}} \right|}^2} + {c_{22}}{{\left| {{\Delta _2}} \right|}^2}} \right)}^2} + c_{12}^2{{\left| {{\Delta _1}} \right|}^2}{{\left| {{\Delta _2}} \right|}^2}} \right], \hfill \\ 
\end{gathered} 
\end{equation}
\end{widetext}

\section{Phase diagram} 
\begin{figure}
\includegraphics[width=0.99\columnwidth]{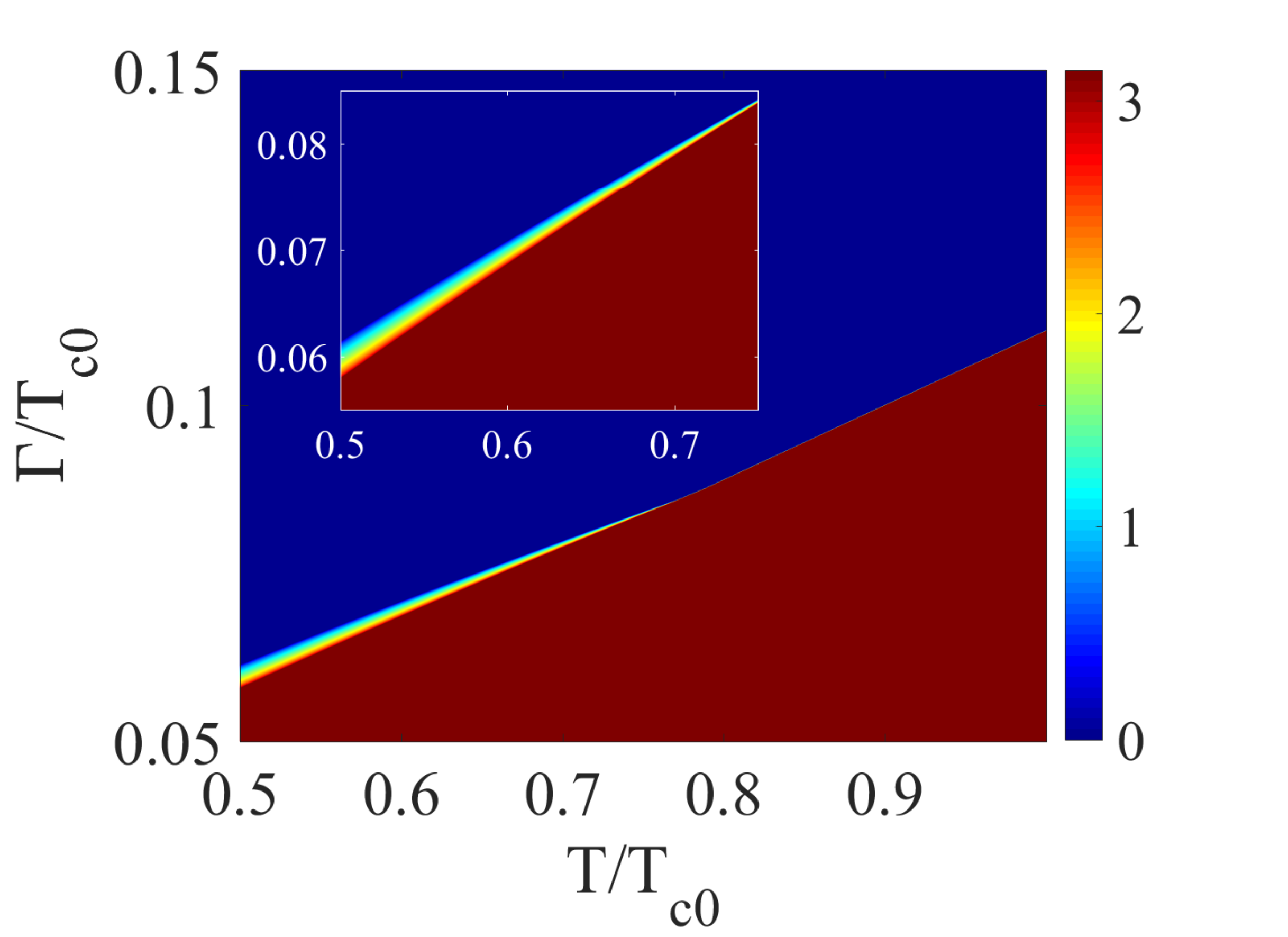}
\caption{Phase diagram of the ground state for a dirty two-band superconductor determining the phase difference $\phi$ as a function of interband scattering rate $\Gamma$ and temperature $T$ (normalized for critical temperature $T_{c0}$ of a clean two-band superconductor) with the set of intra- and interband constants $\lambda_{11}=0.35$, $\lambda_{22}=0.347$,  $\lambda_{12}=\lambda_{21}=-0.01$. For the sake of clarity the zoom of the BTRS domain is shown in the inset.} 
\label{PhD1}
\end{figure}

By solving Eq. (\ref{phi_homo_sol2}) for $\phi$ in the BTRS state one can determine its domain of stability as a function of temperature and interband scattering rate $\Gamma$ in the equilibrium phase when $q=0$, i.e. without a magnetic field. Such case is the initial point for the demonstration of the magneto-topological-induced transitions of the order parameter.
To construct the phase diagram we choose the first set of the expressions for the order parameter moduli as given by Eqs (\ref{OP1_sol1}) and (\ref{OP2_sol1}) and substitute them into Eq. (\ref{phi_homo_sol2}). Based on the microscopic expressions for the coefficients provided in Appendix \ref{sec:A} we show the boundary line of the BTRS state for the intraband $\lambda_{11}=0.35$, $\lambda_{22}=0.347$ and for weak repulsive interband interaction constants $\lambda_{12}=\lambda_{21}=-0.01$ (Fig. \ref{PhD1}).

The narrow region in Fig. \ref{PhD1} corresponds to the BTRS state with $s_{\pm}+is_{++}$ symmetry, while the red and blue regions indicate the emergence of $s_{\pm}$ and $s_{++}$ respectively. We point out that the lower bound of the temperature interval in the phase diagram shown in Fig. \ref{PhD1} may be out of range of the applicability of the GL theory for a dirty two-band superconductor. Thus, one has to apply the microscopic theory for the description in the whole temperature range \cite{Babaev_PD}. Nevertheless, as we will see below this does not affect significantly our conclusions. Moreover, for a given value of the interband scattering rate we choose the temperature in such a way that it is sufficiently close to $T_c$ to obey our phenomenological model calculations (see details in Appendix \ref{sec:C}). 

It should be noted that according to numerical calculations the second set of expressions for the order parameter moduli Eqs (\ref{OP1_sol2}) and (\ref{OP2_sol2}) leads to a similar phase diagram in Fig. \ref{PhD1} with the BTRS domain slightly shifted to larger values of $\Gamma$. In the following we will use the phase diagram based on Eqs. (\ref{OP1_sol1}) and (\ref{OP2_sol1}) since the corresponding solution exhibits the lower energy as shown (discussed) below. 

Finally, the borders of the BTRS domain are determined by the stability conditions deduced from the positive-definite of the determinant of the Hessian matrix that is composed by the second derivatives of the Gibbs free energy with respect to the phase difference and the order parameter moduli.

\section{Magneto-topological transitions}
Now we proceed to the main outcome of our paper. We demonstrate that the application of the magnetic field can lead to competing superconducting configurations marked by a change of the amplitude or the phase of the superconducting order parameter. As a hallmark of the magneto-topological scenario, we find periodic transitions as a function of the magnetic flux.
To illustrate the main outcomes, we choose a representative set of parameters for which the phase diagram has been determine in the equilibrium state (Fig.\ \ref{PhD1}). The temperature and the corresponding value of $\Gamma$ are chosen in the region of the parameters space associated to the BTRS state, where the ``width'' of this region is not vanishing. To comply such a condition we assume that $T=0.7\,T_{c0}$ and $\Gamma=0.07982 \,T_{c0}$. For the given value of $\Gamma$ the critical temperature of a two-band superconductor is approximately $T_c=0.85\,T_{c0}$ as it can be evaluated from the microscopic calculations (see details of the derivation in Appendix \ref{sec:C} and the figure \ref{Tc_vs_Gamma} therein).
\begin{figure}
\includegraphics[width=0.99\columnwidth]{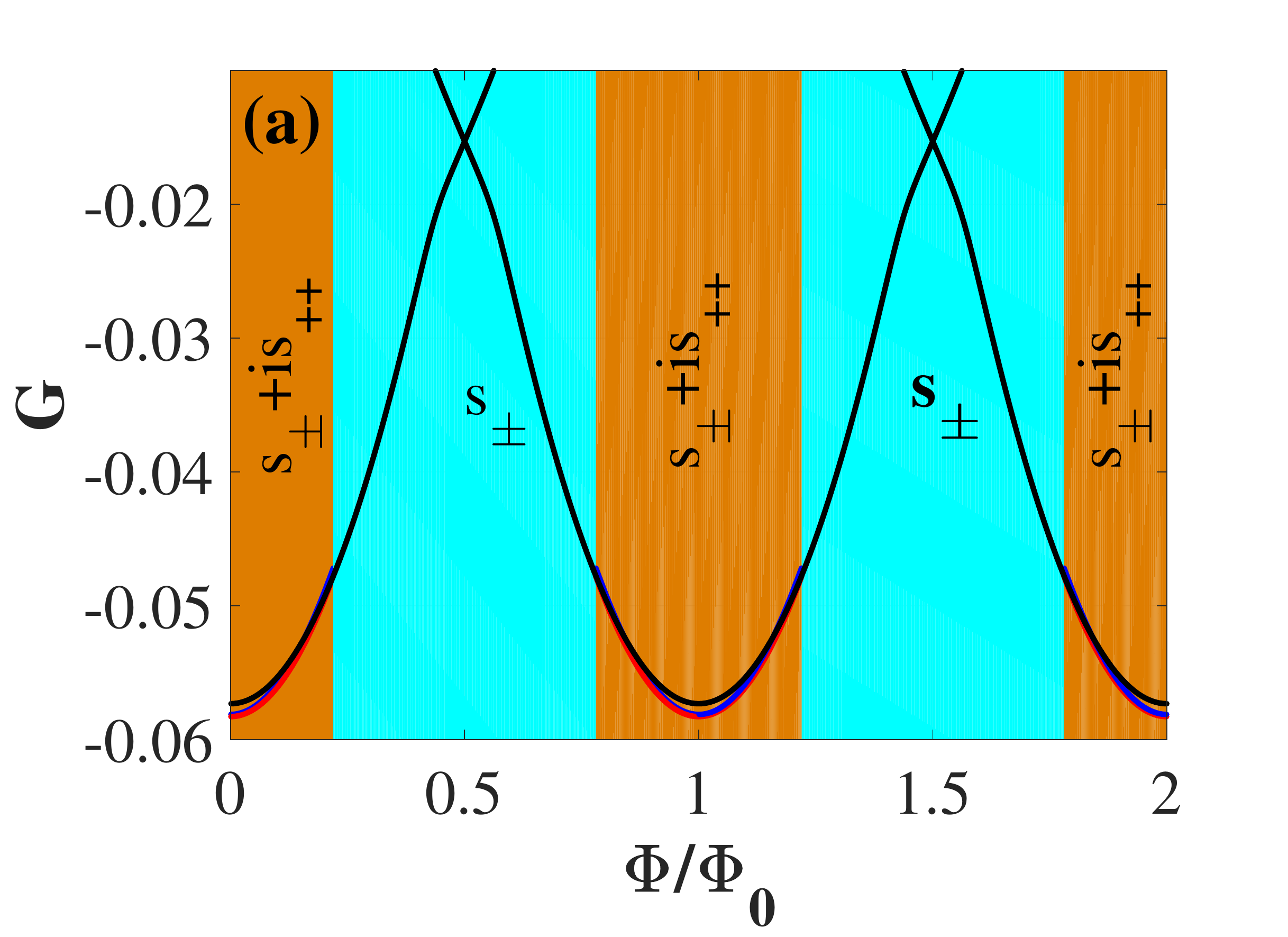}
\includegraphics[width=0.99\columnwidth]{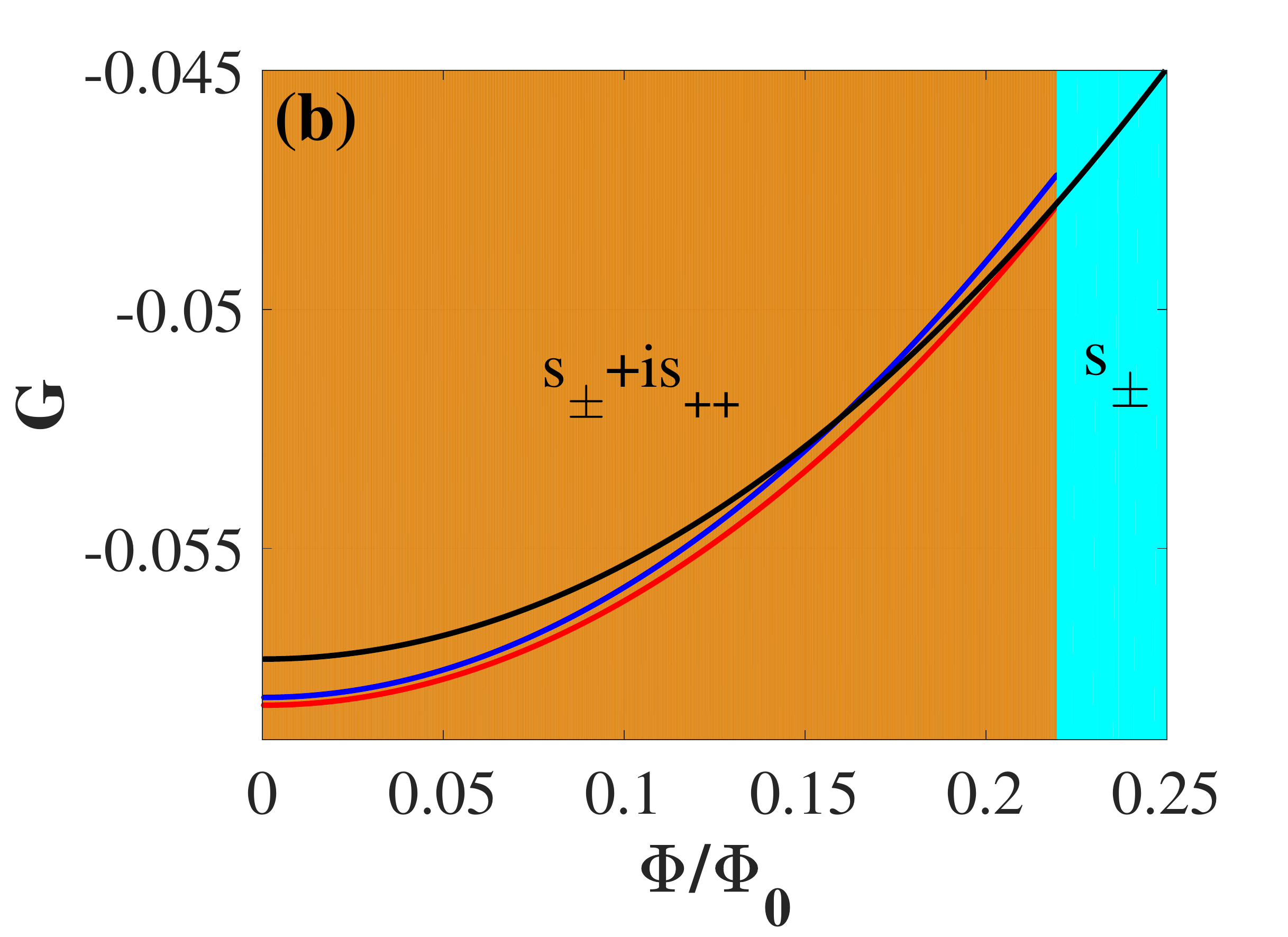}
\caption{(a) The Gibbs free energy of a dirty two-band superconducting cylinder in units of ${N_1}T_{c0}^2{V_s}$ with $\lambda_{11}=0.35$, $\lambda_{22}=0.347$, $\lambda_{12}=\lambda_{21}=-0.01$ and $\Gamma=0.07982/T_{c0}$ for two splitting BTRS states (red and blue lines correspond to solutions with different amplitude of the superconducting order parameter) and for the non-BTRS state (black line). Orange and cyan regions separate domains with different pairing symmetries. (b) Zoom of the phase diagram for values of the magnetic flux associated to the first sequence of transitions. The ratio of diffusion coefficients $D_2/D_1=2$.} 
\label{Energy_plot}
\end{figure}

Then, we compare the Gibbs free energy of BTRS and non-BTRS states with $s_{\pm}+is_{++}$ and $s_{\pm}$ symmetry, respectively, as a function of applied magnetic flux when $q \ne 0$. We perform numerical solutions for $|\Delta_1|$ and $|\Delta_2|$ on a dependence of $q$ are then substituted into expressions for $G$ of the non-BTRS state with $s_{\pm}$ pairing symmetry Eq. (\ref{Gibbs_final}) and of the BTRS state with $s_{\pm}+is_{++}$ symmetry Eq. (\ref{Gibbs_final_new}). The behavior of these energies is shown in Figure \ref{Energy_plot}. One can see that $G$ of the BTRS state (blue and red lines) either crosses (blue line) the curve of the $G$ for the non-BTRS state (black line) or just touches it (red line). In the latter case the intersection occurs at the boundary of the stability region of the BTRS state (see the zoom of Fig. \ref{Energy_plot}). Hence, we demonstrate that a periodic oscillation from $s_{\pm}+is_{++}$ to $s_{\pm}$ and vice versa is acheived in the doubly connected topology due to the magnetic field. 

At first glance it may seem somewhat surprising to have two different stable energy states in the BTRS domain. However, firstly we recall the existence of two stable solutions for $|\Delta_i|$ in the equilibrium state as given by Eqs.\ (\ref{OP1_sol1}-\ref{OP2_sol2}), and as a consequence of them there are two distinct boundary lines. Secondly, the BTRS is a superposition of two different superconducting components. They behave like a doublet and the presence of interband impurities acts as an effective magnetic field thus inducing in-equivalent configurations. 
This scenario is based on the assumption that the impurity scattering is weak enough not to induce a transition to an $s_{++}$ state in the bulk. To the best of our knowledge this issue has not yet been addressed for the case of weak repulsive interband couplings in the literature. In this context, one has to refer to other studies which have been developed within the framework of Eliashberg theory \cite{Efremov2011,Efremov2017}. For our approach, one can use the $T_c$-value obtained for vanishing repulsive interband couplings and very small attractive interactions yielding 0.817735 $T_{c0}$ in the weak coupling case and the intraband parameters considered above
(see Eqs. (\ref{estimation_Tc}) and (\ref{estimation_Tc_new}) in Appendix \ref{sec:D}). Such value is still well below the point (blue) corresponding to 0.8485 $T_{c0}$ as shown in Fig. \ref{Tc_vs_Gamma} in Appendix \ref{sec:C}. The analysis for realistic impurity couplings and configurations is left for future investigations. 

Our numerical calculations admit the onset of oscillations between the $s_{\pm}+is_{++}$ and $s_{++}$ type symmetries of the order parameter for large values of $\Gamma$, at the upper border of the BTRS domain (see Fig. \ref{PhD1}). However, within the microscopic consideration it has been shown already that for the strong inter-band scattering effect (large values of $\Gamma$) a two-band superconductor can behave as an effective one-band dirty superconductor \cite{Ng2009}. Since the magneto-topological scenario is introduced within the phenomenological approach we focus on transitions from an $s_{\pm}+is_{++}$ state to an $s_{\pm}$ state and vice versa, which occur for small values of $\Gamma$.

\section{Discussion}  
We argue that the unveiled magneto-topological transitions are not only relevant for multiband superconductors but also for artificially engineered systems with competing 0- and $\pi$-Josephson couplings \cite{Lin,Guarcello}. Moreover, while the results have been demonstrated for the case of a cylinder, they can be directly extended to other superconducting loops having an Euler characteristic that is zero like for the torus and the M\"obius strip. In the latter case, one may expect a richer scenario of transitions from chiral to non-chiral configurations thus augmenting the manifestations of the magneto-topological-induced scenario. It should be noted that this topological requirement of zero Euler characteristic is essential for the quantization of phases of the multi-component order parameter.

Let us point out that inhomogeneous states, like phase solitons due to additional degrees of freedom of the multi-component order parameter, have significantly higher energies compared to the homogeneous states addressed here \cite{Yerin1, Yerin2, Yerin6}. Therefore, we excluded them from the present study. 

Although the analysis has been performed for a two-component superconductor, the form of Eq.\ (\ref{Gibbs_final_new}) suggests another generalization of our results. Indeed, Eq. (\ref{Gibbs_final_new}) formally reminds the structure of the GL energy in the case of an FFLO state due to its fourth-degree polynomial in terms of $q$ \cite{Buzdin1, Samokhin1}. This analogy indicates the possibility of having magneto-topological-induced transitions for the FFLO state in conventional superconductor-ferromagnetic (S-F) heterostructures with the doubly-connected geometry \cite{Mironov}. There, instead of inducing transitions by means of temperature or material parameters (e.g. thickness of the S or F layers, conductivity, etc.) one can manipulate homogeneous non-FFLO and FFLO states by means of the magnetic flux.

Another interesting perspective is to consider a dynamical manipulation of the chiral and time-reversal symmetric states. 
It is known that ultrafast light allows to control different states of matter, also encompassing the phenomenon of superconductivity. For instance by light pulse, one can cause a superconducting state to appear for a short period even at temperatures that are higher than $T_c$ \cite{Cavalleri1, Cavalleri2, Cavalleri3, Cavalleri4, Isoyama}. Here, we envision the possibility of inducing either amplitude or phase oscillations by employing a time dependent perturbation which can couple the $s_{\pm}+is_{++}$ and $s_{\pm}$ superconducting configurations. Thus, we argue that a sort of \textit{dynamical chiral superconductivity} can be obtained by suitably using a combination of static and time dependent electromagnetic fields.

From an experimental point of view, the periodic transitions of the superconducting phases can be detected by probing the current-induced magnetic flux response. Since the supercurrent $j$ in the loop is given by $j \sim \partial G/\partial q$ it directly follows that a magnetic flux should induce jumps in the current density.

\section{Conclusions}
We have demonstrated that a superconducting phase with BTRS arising from a phase frustration between 0- and $\pi$- pairing will undergo a transition into a time-reversal symmetric state by applying a magnetic field in a non-simple connected geometry. This finding can be qualitatively understood by observing that the interband phase frustration can be released by the presence of the magnetic flux because the magnetic vector potential directly affects the relative phase of the superconducting components. Then, a time-reversal symmetric configuration dominated by one of the two pairing channels becomes energetically favorable. In this context, one can also expect that a transition from $s_{\pm}+i\,s_{++}$ to $s_{++}$ might emerge in suitable microscopic conditions. The unveiled magneto-topological transitions resembles the case of triangular spin-frustrated systems  where the application of magnetic field leads to a transition from a chiral (non-collinear) spin-state to a collinear one. 
Along this line, we argue that dynamical effects can be exploited for accessing the structure of unconventional superconductors by searching for transitions between chiral states having different amplitudes of the order parameter or from chiral to non-chiral phases.

\begin{acknowledgments}
Y.Y.\ acknowledges support by the CarESS project. We thank D.\ Efremov for discussions.
\end{acknowledgments}

\begin{widetext}
\appendix
\section{GL coefficients}
\label{sec:A}

The coefficients of the GL theory, derived from the microscopic Usadel equations, are defined as follows \cite{Stanev, Corticelli}:
\begin{equation}
\label{a_i}
{a_{ii}} = {N_i}\left( {\frac{{{\lambda _{jj}}}}{{\det {\lambda _{ij}}}} - 2\pi T\sum\limits_{\omega  > 0}^{{\omega _c}} {\frac{{\omega  + {\Gamma _{ij}}}}{{\omega \left( {\omega  + {\Gamma _{ij}} + {\Gamma _{ji}}} \right)}}} } \right) = {N_i}\left( {\frac{{{\lambda _{jj}}}}{{\det {\lambda _{ij}}}} - \frac{1}{\lambda } + \ln \left( {\frac{T}{{{T_c}}}} \right) + \psi \left( {\frac{1}{2} + \frac{\Gamma }{{\pi T}}} \right) - \psi \left( {\frac{1}{2}} \right)} \right),
\end{equation}
\begin{equation}
\label{a_12}
{a_{ij}} =  - {N_i}\left( {\frac{{{\lambda _{ij}}}}{{\det {\lambda _{ij}}}} + 2\pi T\sum\limits_{\omega  > 0}^{{\omega _c}} {\frac{{{\Gamma _{ij}}}}{{\omega \left( {\omega  + {\Gamma _{ij}} + {\Gamma _{ji}}} \right)}}} }, \right)
\end{equation}
\begin{equation}
\label{b_i}
{b_{ii}} = {N_i}\pi T\sum\limits_{\omega  > 0}^{{\omega _c}} {\frac{{{{\left( {\omega  + {\Gamma _{ji}}} \right)}^4}}}{{{\omega ^3}{{\left( {\omega  + {\Gamma _{ij}} + {\Gamma _{ji}}} \right)}^4}}}}  + {N_i}\pi T\sum\limits_{\omega  > 0}^{{\omega _c}} {\frac{{{\Gamma _{ij}}\left( {\omega  + {\Gamma _{ji}}} \right)\left( {{\omega ^2} + 3\omega {\Gamma _{ji}} + \Gamma _{ji}^2} \right)}}{{{\omega ^3}{{\left( {\omega  + {\Gamma _{ij}} + {\Gamma _{ji}}} \right)}^4}}}},
\end{equation}
\begin{equation}
\label{b_12}
{b_{ij}} =  - {N_i}\pi T\sum\limits_{\omega  > 0}^{{\omega _c}} {\frac{{{\Gamma _{ij}}{\omega ^3}}}{{{\omega ^3}{{\left( {\omega  + {\Gamma _{ij}} + {\Gamma _{ji}}} \right)}^4}}}}  + {N_i}\pi T\sum\limits_{\omega  > 0}^{{\omega _c}} {\frac{{{\Gamma _{ij}}\left( {{\Gamma _{ij}} + {\Gamma _{ji}}} \right)\left( {{\Gamma _{ji}}\left( {\omega  + 2{\Gamma _{ij}}} \right) + \omega {\Gamma _{ij}}} \right)}}{{{\omega ^3}{{\left( {\omega  + {\Gamma _{ij}} + {\Gamma _{ji}}} \right)}^4}}}},
\end{equation}
\begin{equation}
\label{c_i}
{c_{ii}} = {N_i}\pi T\sum\limits_{\omega  > 0}^{{\omega _c}} {\frac{{{\Gamma _{ij}}\left( {\omega  + {\Gamma _{ji}}} \right)\left( {{\omega ^2} + \left( {\omega  + {\Gamma _{ji}}} \right)\left( {{\Gamma _{ij}} + {\Gamma _{ji}}} \right)} \right)}}{{{\omega ^3}{{\left( {\omega  + {\Gamma _{ij}} + {\Gamma _{ji}}} \right)}^4}}}},
\end{equation}
\begin{equation}
\label{c_12}
{c_{ij}} = {N_i}\pi T\sum\limits_{\omega  > 0}^{{\omega _c}} {\frac{{{\Gamma _{ij}}\left( {\omega  + {\Gamma _{ji}}} \right)\left( {\omega  + {\Gamma _{ji}}} \right)\left( {{\Gamma _{ij}} + {\Gamma _{ji}}} \right)}}{{{\omega ^3}{{\left( {\omega  + {\Gamma _{ij}} + {\Gamma _{ji}}} \right)}^4}}}},
\end{equation}
\begin{equation}
\label{k_i}
{k_{ii}} = 2{N_i}\pi T\sum\limits_{\omega  > 0}^{{\omega _c}} {\frac{{{D_i}{{\left( {\omega  + {\Gamma _{ji}}} \right)}^2} + {\Gamma _{ij}}{\Gamma _{ji}}{D_j}}}{{{\omega ^2}{{\left( {\omega  + {\Gamma _{ij}} + {\Gamma _{ji}}} \right)}^2}}}}
\end{equation}
\begin{equation}
\label{k_12}
{k_{ij}} = 2{N_i}{\Gamma _{ij}}\pi T\sum\limits_{\omega  > 0}^{{\omega _c}} {\frac{{{D_i}\left( {\omega  + {\Gamma _{ji}}} \right) + {D_j}\left( {\omega  + {\Gamma _{ij}}} \right)}}{{{\omega ^2}{{\left( {\omega  + {\Gamma _{ij}} + {\Gamma _{ji}}} \right)}^2}}}},
\end{equation}
where $\omega=(2n+1)\pi T$  are Matsubara frequencies, $\omega_c$ is the cut-off frequency,  $N_i$ are the densities of states at the Fermi level, $\lambda_{ij}$  and $\Gamma_{ij}$ are coupling constants and interband scattering rates that characterize the strength of the interband impurities, $D_i$ are diffusion coefficients. For the sake of simplicity and without loss of generality we put $\lambda_{12}=\lambda_{21}$, $\Gamma_{12}=\Gamma_{21}$ and $N_1=N_2$ in the main paper.

In principle, Eqs.\ (\ref{a_12})-(\ref{k_12}) admit exact summation and can be expressed in terms of polygamma functions. However, we do not provide these expression due to their cumbersome forms. 

\section{Diagonalization of the Gibbs free energy and the derivation of main equations}
\label{sec:B}

To diagonalize the functional given by Eq. (\ref{GL_general}) we introduce new functional variables: the phase difference $\phi$ and the weighted average phase $\theta$ \cite{Yerin4}
\begin{equation}
\label{diag_initial}
\left\{ {\begin{array}{*{20}{c}}
  {{\chi _1} - {\chi _2} = \phi ,} \\ 
  {{l_1}{\chi _1} + {l_2}{\chi _2} = \theta,} 
\end{array}} \right.
\end{equation}
where $l_1$ and $l_2$ are some coefficients to be determined below.

To determine the ratio between the new and old functional variables entering Eqs. (\ref{diag_initial}) must be solved
\begin{equation}
\label{diag_initial_new}
\left\{ {\begin{array}{*{20}{c}}
  {\nabla {\chi _1} = \frac{1}{{{l_1} + {l_2}}}\nabla \theta  + \frac{{{l_2}}}{{{l_1} + {l_2}}}\nabla \phi ,} \\ 
  {\nabla {\chi _2} = \frac{1}{{{l_1} + {l_2}}}\nabla \theta  - \frac{{{l_1}}}{{{l_1} + {l_2}}}\nabla \phi .} 
\end{array}} \right.
\end{equation}

After the substitution of Eqs.\ (\ref{diag_initial_new}) the expressions for the partial and interband components of the Gibbs free energy entering Eqs. (\ref{GL_general1})-(\ref{GL_general3}) transform to
\begin{equation}
\label{GL_diag1}
\begin{gathered}
  {F_1} = \int {\left[ {{a_{11}}{{\left| {{\Delta _1}} \right|}^2} + \frac{1}{2}{b_{11}}{{\left| {{\Delta _1}} \right|}^4} + \frac{1}{2}{k_{11}}{\hbar ^2}{{\left| {{\Delta _1}} \right|}^2}{{\left( {\frac{1}{{{l_1} + {l_2}}}\nabla \theta  + \frac{{{l_2}}}{{{l_1} + {l_2}}}\nabla \phi  - \frac{{2e}}{{c\hbar }}{\mathbf{A}}} \right)}^2}} \right]} {d^3}{\mathbf{r}}, \hfill \\ 
\end{gathered}
\end{equation}
\begin{equation}
\label{GL_diag2}
\begin{gathered}
  {F_2} =\int {\left[ {{a_{22}}{{\left| {{\Delta _2}} \right|}^2} + \frac{1}{2}{b_{22}}{{\left| {{\Delta _2}} \right|}^4} + \frac{1}{2}{k_{22}}{\hbar ^2}{{\left| {{\Delta _2}} \right|}^2}{{\left( {\frac{1}{{{l_1} + {l_2}}}\nabla \theta  - \frac{{{l_1}}}{{{l_1} + {l_2}}}\nabla \phi  - \frac{{2e}}{{c\hbar }}{\mathbf{A}}} \right)}^2}} \right]} {d^3}{\mathbf{r}}, \hfill \\ 
\end{gathered}
\end{equation}
\begin{equation}
\label{GL_diag3}
\begin{gathered}
  {F_{12}} = \int {\left[ {2\left( {{a_{12}}\left| {{\Delta _1}} \right|\left| {{\Delta _2}} \right| + {c_{11}}{{\left| {{\Delta _1}} \right|}^3}\left| {{\Delta _2}} \right| + {c_{22}}\left| {{\Delta _1}} \right|{{\left| {{\Delta _2}} \right|}^3}} \right)\cos \phi  + {c_{12}}{{\left| {{\Delta _1}} \right|}^2}{{\left| {{\Delta _2}} \right|}^2}\cos 2\phi  + {b_{12}}{{\left| {{\Delta _1}} \right|}^2}{{\left| {{\Delta _2}} \right|}^2}} \right.}  \hfill \\
  \left. { + {k_{12}}{\hbar ^2}\left| {{\Delta _1}} \right|\left| {{\Delta _2}} \right|\left( {\frac{1}{{{l_1} + {l_2}}}\nabla \theta  + \frac{{{c_2}}}{{{c_1} + {c_2}}}\nabla \phi  - \frac{{2e}}{{c\hbar }}{\mathbf{A}}} \right)\left( {\frac{1}{{{l_1} + {l_2}}}\nabla \theta  - \frac{{{l_1}}}{{{l_1} + {l_2}}}\nabla \phi  - \frac{{2e}}{{c\hbar }}{\mathbf{A}}} \right)\cos \phi } \right]{d^3}{\mathbf{r}}. \hfill \\ 
\end{gathered}
\end{equation}

Putting  ${l_1} + {l_2} = 1$ and setting zero the coefficient with the product term  $\nabla \theta  \cdot \nabla \phi $ one can obtain explicit expressions for the coefficients $l_1$ and $l_2$
\begin{equation}
\label{l1l2}
\begin{gathered}
  {l_1} = \frac{{{k_{11}}{{\left| {{\Delta _1}} \right|}^2} + {k_{12}}\left| {{\Delta _1}} \right|\left| {{\Delta _2}} \right|\cos \phi }}{{{k_{11}}{{\left| {{\Delta _1}} \right|}^2} + {k_{22}}{{\left| {{\Delta _2}} \right|}^2} + 2{k_{12}}\left| {{\Delta _1}} \right|\left| {{\Delta _2}} \right|\cos \phi }},{\text{ }} \hfill \\
  {l_2} = \frac{{{k_{22}}{{\left| {{\Delta _2}} \right|}^2} + {k_{12}}\left| {{\Delta _1}} \right|\left| {{\Delta _2}} \right|\cos \phi }}{{{k_{11}}{{\left| {{\Delta _1}} \right|}^2} + {k_{22}}{{\left| {{\Delta _2}} \right|}^2} + 2{k_{12}}\left| {{\Delta _1}} \right|\left| {{\Delta _2}} \right|\cos \phi }}. \hfill \\ 
\end{gathered} 
\end{equation}

Irrespective of a specific topology of a system under consideration after the diagonalization procedure for the density of the Gibbs free energy, $\mathbb{G}$ can be rewritten in the compact form
\begin{equation}
\label{GL_diag}
\mathbb{G} = {\mathbb{F}_0}+A{\left( {\nabla \theta  - \frac{{2e}}{{c\hbar }}{\mathbf{A}}} \right)^2} + B{\left( {\nabla \phi } \right)^2} + C\cos \phi  + D\cos 2\phi 
\end{equation}
where 
\begin{equation}
\label{GL_F0}
{\mathbb{F}_0} = {a_{11}}{\left| {{\Delta _1}} \right|^2} + \frac{1}{2}{b_{11}}{\left| {{\Delta _1}} \right|^4} + {a_{22}}{\left| {{\Delta _2}} \right|^2} + \frac{1}{2}{b_{22}}{\left| {{\Delta _2}} \right|^4}+{b_{12}}{\left| {{\Delta _1}} \right|^2}{\left| {{\Delta _2}} \right|^2},
\end{equation}
\begin{equation}
\label{GL_A}
A = \left( {\frac{1}{2}{k_{11}}{{\left| {{\Delta _1}} \right|}^2} + \frac{1}{2}{k_{22}}{{\left| {{\Delta _2}} \right|}^2} + {k_{12}}\left| {{\Delta _1}} \right|\left| {{\Delta _2}} \right|\cos \phi } \right){\hbar ^2},
\end{equation}
\begin{equation}
\label{GL_B}
B = \left( {\frac{1}{2}l_2^2{k_{11}}{{\left| {{\Delta _1}} \right|}^2} + \frac{1}{2}l_1^2{k_{22}}{{\left| {{\Delta _2}} \right|}^2} - {l_1}{l_2}{k_{12}}\left| {{\Delta _1}} \right|\left| {{\Delta _2}} \right|\cos \phi } \right){\hbar ^2},
\end{equation}
\begin{equation}
\label{GL_C}
C = 2\left( {{a_{12}}\left| {{\Delta _1}} \right|\left| {{\Delta _2}} \right| + {c_{11}}{{\left| {{\Delta _1}} \right|}^3}\left| {{\Delta _2}} \right| + {c_{22}}\left| {{\Delta _1}} \right|{{\left| {{\Delta _2}} \right|}^3}} \right),
\end{equation}
\begin{equation}
\label{GL_D}
D = {c_{12}}{\left| {{\Delta _1}} \right|^2}{\left| {{\Delta _2}} \right|^2}.
\end{equation}

The variational procedure applied to Eq.\ (\ref{GL_diag}) yields the Euler-Lagrange equations for the two phase variables $\theta$ and $\phi$
\begin{equation}
\label{EL_GL}
\left\{ {\begin{array}{*{20}{c}}
  { - \left( {{k_{11}}{{\left| {{\Delta _1}} \right|}^2} + {k_{22}}{{\left| {{\Delta _2}} \right|}^2} + 2{k_{12}}\left| {{\Delta _1}} \right|\left| {{\Delta _2}} \right|\cos \phi } \right){\nabla ^2}\left( {\theta  - \frac{{2e}}{{c\hbar }}{\mathbf{A}}} \right) + 2{k_{12}}\left| {{\Delta _1}} \right|\left| {{\Delta _2}} \right|\sin \phi \nabla \left( {\theta  - \frac{{2e}}{{c\hbar }}{\mathbf{A}}} \right)\nabla \phi  = 0,} \\ 
  \begin{gathered}
  \frac{{\left( {{k_{11}}{k_{22}} - k_{12}^2{{\cos }^2}\phi } \right){{\left| {{\Delta _1}} \right|}^2}{{\left| {{\Delta _2}} \right|}^2}}}{{{k_{11}}{{\left| {{\Delta _1}} \right|}^2} + {k_{22}}{{\left| {{\Delta _2}} \right|}^2} + 2{k_{12}}\left| {{\Delta _1}} \right|\left| {{\Delta _2}} \right|\cos \phi }}{\hbar ^2}{\nabla ^2}\phi  + {k_{12}}\left| {{\Delta _1}} \right|\left| {{\Delta _2}} \right|\sin \phi {\left( {\hbar \nabla \left( {\theta  - \frac{{2e}}{{c\hbar }}{\mathbf{A}}} \right)} \right)^2} \hfill \\
   - \frac{{{k_{12}}\left( {{k_{11}}\left| {{\Delta _1}} \right| + \left| {{\Delta _2}} \right|\cos \phi } \right)\left( {{k_{22}}\left| {{\Delta _2}} \right| + \left| {{\Delta _1}} \right|\cos \phi } \right){{\left| {{\Delta _1}} \right|}^2}{{\left| {{\Delta _2}} \right|}^2}\sin \phi }}{{{{\left( {{k_{11}}{{\left| {{\Delta _1}} \right|}^2} + {k_{22}}{{\left| {{\Delta _2}} \right|}^2} + 2{k_{12}}\left| {{\Delta _1}} \right|\left| {{\Delta _2}} \right|\cos \phi } \right)}^2}}}{\left( {\hbar \nabla \phi } \right)^2} \hfill \\
   + 2\left( {{a_{12}}\left| {{\Delta _1}} \right|\left| {{\Delta _2}} \right| + {c_{11}}{{\left| {{\Delta _1}} \right|}^3}\left| {{\Delta _2}} \right| + {c_{22}}\left| {{\Delta _1}} \right|{{\left| {{\Delta _2}} \right|}^3}} \right)\sin \phi  + 2{c_{12}}{\left| {{\Delta _1}} \right|^2}{\left| {{\Delta _2}} \right|^2}\sin 2\phi  = 0. \hfill \\ 
\end{gathered}  
\end{array}} \right.
\end{equation}

The first integrals of Eq. (\ref{EL_GL}) take the form
\begin{equation}
\label{EL_GL_1st}
\left\{ {\begin{array}{*{20}{c}}
  {\left( {{k_{11}}{{\left| {{\Delta _1}} \right|}^2} + {k_{22}}{{\left| {{\Delta _2}} \right|}^2} + 2{k_{12}}\left| {{\Delta _1}} \right|\left| {{\Delta _2}} \right|\cos \phi } \right)\nabla \left( {\theta  - \frac{{2e}}{{c\hbar }}{\mathbf{A}}} \right) = {K_1},} \\ 
  \begin{gathered}
  \left( {{k_{11}}{{\left| {{\Delta _1}} \right|}^2} + {k_{22}}{{\left| {{\Delta _2}} \right|}^2} + 2{k_{12}}\left| {{\Delta _1}} \right|\left| {{\Delta _2}} \right|\cos \phi } \right){\left( {\hbar \nabla \left( {\theta  - \frac{{2e}}{{c\hbar }}{\mathbf{A}}} \right)} \right)^2} + \frac{{\left( {{k_{11}}{k_{22}} - k_{12}^2{{\cos }^2}\phi } \right){{\left| {{\Delta _1}} \right|}^2}{{\left| {{\Delta _2}} \right|}^2}}}{{{k_{11}}{{\left| {{\Delta _1}} \right|}^2} + {k_{22}}{{\left| {{\Delta _2}} \right|}^2} + 2{k_{12}}\left| {{\Delta _1}} \right|\left| {{\Delta _2}} \right|\cos \phi }}{\left( {\hbar \nabla \phi } \right)^2} \hfill \\
   - 4\left( {{a_{12}}\left| {{\Delta _1}} \right|\left| {{\Delta _2}} \right| + {c_{11}}{{\left| {{\Delta _1}} \right|}^3}\left| {{\Delta _2}} \right| + {c_{22}}\left| {{\Delta _1}} \right|{{\left| {{\Delta _2}} \right|}^3}} \right)\cos \phi  - 2{c_{12}}{\left| {{\Delta _1}} \right|^2}{\left| {{\Delta _2}} \right|^2}\cos 2\phi  = {K_2}, \hfill \\ 
\end{gathered}  
\end{array}} \right.
\end{equation}
where $K_1$ and $K_2$ are constants of the integration.

The first equation of the system Eq.\ (\ref{EL_GL_1st}) allows to express the gradient $\nabla \left( {\theta  - \frac{{2e}}{{c\hbar }}{\mathbf{A}}} \right)$  as a function of the second variable $\phi$ and and to substitute it into the second equation, thereby obtaining a nonlinear differential equation of the first order for $\phi$
\begin{equation}
\label{phi_general}
\begin{gathered}
  \frac{{{\hbar ^2}K_1^2}}{{{k_{11}}{{\left| {{\Delta _1}} \right|}^2} + {k_{22}}{{\left| {{\Delta _2}} \right|}^2} + 2{k_{12}}\left| {{\Delta _1}} \right|\left| {{\Delta _2}} \right|\cos \phi }} + \frac{\hbar^2{\left( {{k_{11}}{k_{22}} - k_{12}^2{{\cos }^2}\phi } \right){{\left| {{\Delta _1}} \right|}^2}{{\left| {{\Delta _2}} \right|}^2}}}{{{k_{11}}{{\left| {{\Delta _1}} \right|}^2} + {k_{22}}{{\left| {{\Delta _2}} \right|}^2} + 2{k_{12}}\left| {{\Delta _1}} \right|\left| {{\Delta _2}} \right|\cos \phi }}{\left( { \nabla \phi } \right)^2} \hfill \\
   - 4\left( {{a_{12}}\left| {{\Delta _1}} \right|\left| {{\Delta _2}} \right| + {c_{11}}{{\left| {{\Delta _1}} \right|}^3}\left| {{\Delta _2}} \right| + {c_{22}}\left| {{\Delta _1}} \right|{{\left| {{\Delta _2}} \right|}^3}} \right)\cos \phi  - 2{c_{12}}{\left| {{\Delta _1}} \right|^2}{\left| {{\Delta _2}} \right|^2}\cos 2\phi  = {K_2}. \hfill \\ 
\end{gathered}
\end{equation}

Eq. (\ref{phi_general}) provides an important tool for the study of all possible inhomogeneous solutions like FFLO state, phase solitons and other possible exotic phases for dirty two-band superconductors \cite{Yerin3, Tanaka2002, Gurevich2003, Lin2, Vakaryuk, Babaev5, Arisawa, Samokhin2, Vodolazov, Yerin5, Machida1, Ptok, Machida2}. We would like to note that the theoretical prediction of phase solitons has been obtained in Ref. \onlinecite{Tanaka2002} for an open one-dimensional geometry within the sine-Gordon model assuming the characteristic kink solution. There, phase soliton solutions for a ring are shortly discussed assuming a single winding number only. Hereafter we consider the case where soliton solutions are parametrized by two winding numbers corresponding to phases of the two-component order parameter.

Being topological defects phase solitons are forbidden in the bulk due to divergent total energy in the spatially unlimited case, but they can have finite energy in special doubly connected topologies like in a thin-walled cylinder. In this case introducing cylindrical coordinates Eqs. (\ref{EL_GL}) must be supplemented by boundary conditions for each phase $\chi_i$ of the order parameter:
\begin{equation}
\label{quantization_condition}
\oint\limits_C  {\nabla {\chi _i} \cdot d{\mathbf{l}}}  = 2\pi {N_i},
\end{equation}
where $C$ is an arbitrary closed contour that lies inside the wall of the cylinder and encircles the opening and $\; N_{i} =0,\pm 1,\pm 2,...$ are winding numbers. As the result of the symmetry of the problem and the continuity conditions this gives
\begin{equation}
\label{bc_chi12}
\begin{gathered}
  {\left. {{\chi _{1,2}}} \right|_{\varphi  = 2\pi }} - {\left. {{\chi _{1,2}}} \right|_{\varphi  = 0}} = 2\pi {N_{1,2}}, \hfill \\
  {\left. {\frac{{d{\chi _{1,2}}}}{{d\varphi }}} \right|_{\varphi  = 0}} = {\left. {\frac{{d{\chi _{1,2}}}}{{d\varphi }}} \right|_{\varphi  = 2\pi }},\quad {N_{1,2}} = 0, \pm 1, \pm 2,... \hfill \\ 
\end{gathered} 
\end{equation}
with the corresponding boundary conditions for the phase variables $\theta$
\begin{equation}
\label{bc_theta}
\begin{gathered}
  {\left. \theta  \right|_{\varphi  = 2\pi }} - {\left. \theta  \right|_{\varphi  = 0}} = 2\pi \left( {{l_1}{N_1} + {l_2}{N_2}} \right),\quad  \hfill \\
  {\left. {\frac{{d\theta }}{{d\varphi }}} \right|_{\varphi  = 0}} = {\left. {\frac{{d\theta }}{{d\varphi }}} \right|_{\varphi  = 2\pi }},\quad {N_{1,2}} = 0, \pm 1, \pm 2,... \hfill \\ 
\end{gathered}
\end{equation}
and $\phi$
\begin{equation}
\label{bc_phi}
\begin{gathered}
  {\left. \phi  \right|_{\varphi  = 2\pi }} - {\left. \phi  \right|_{\varphi  = 0}} = 2\pi n,\quad  \hfill \\
  {\left. {\frac{{d\phi }}{{d\varphi }}} \right|_{\varphi  = 0}} = {\left. {\frac{{d\phi }}{{d\varphi }}} \right|_{\varphi  = 2\pi }},\quad n = {N_1} - {N_2} = 0, \pm 1, \pm 2,... \hfill \\ 
\end{gathered}
\end{equation}
where $\varphi$ is the polar angle.

Since we are interested in a homogeneous state of the system $N_1=N_2$, i.e.\ ignoring boundary effects of the tube,  Eqs. (\ref{EL_GL}) can be significantly simplified
\begin{equation}
\label{theta_phi_homo}
\left\{ {\begin{array}{*{20}{c}}
  {\frac{{{\partial ^2}\theta }}{{\partial {\varphi ^2}}} = 0,} \\ 
  \begin{gathered}
  {k_{12}}\left| {{\Delta _1}} \right|\left| {{\Delta _2}} \right|\sin \phi \frac{{{\hbar ^2}}}{{{R^2}}}{\left( {\frac{{\partial \theta }}{{\partial \varphi }} - \frac{\Phi }{{{\Phi _0}}}} \right)^2} \hfill \\
   + 2\left( {{a_{12}}\left| {{\Delta _1}} \right|\left| {{\Delta _2}} \right| + {c_{11}}{{\left| {{\Delta _1}} \right|}^3}\left| {{\Delta _2}} \right| + {c_{22}}\left| {{\Delta _1}} \right|{{\left| {{\Delta _2}} \right|}^3}} \right)\sin \phi  \hfill \\
   + 2{c_{12}}{\left| {{\Delta _1}} \right|^2}{\left| {{\Delta _2}} \right|^2}\sin 2\phi  = 0. \hfill \\ 
\end{gathered}  
\end{array}} \right.
\end{equation}

The solution of the first equation in the system of Eq. (\ref{theta_phi_homo}) for $\theta$ is represented by a linear function of the winding number $N=N_1=N_2$
\begin{equation}
\label{theta_homo}
\theta(\varphi)=N\varphi + \theta(0).
\end{equation}
In the case of a thin-walled cylinder the Gibbs free energy acquires the form
\begin{equation}
\label{GL_integral_zr}
\begin{gathered}
  \frac{F}{{{V_s}}} = {F_0} + \int\limits_0^{2\pi } {\frac{{d\varphi }}{{2\pi }}\left[ {\left( {\frac{1}{2}{k_{11}}{{\left| {{\Delta _1}} \right|}^2} + \frac{1}{2}{k_{22}}{{\left| {{\Delta _2}} \right|}^2} + {k_{12}}\left| {{\Delta _1}} \right|\left| {{\Delta _2}} \right|\cos \phi } \right)\frac{{{\hbar ^2}}}{{{R^2}}}{{\left( {\frac{{\partial \theta }}{{\partial \varphi }} - \frac{\Phi }{{{\Phi _0}}}} \right)}^2} + } \right.}  \hfill \\
  \left( {\frac{1}{2}l_2^2{k_{11}}{{\left| {{\Delta _1}} \right|}^2} + \frac{1}{2}l_1^2{k_{22}}{{\left| {{\Delta _2}} \right|}^2} - {l_1}{l_2}{k_{12}}\left| {{\Delta _1}} \right|\left| {{\Delta _2}} \right|\cos \phi } \right)\frac{{{\hbar ^2}}}{{{R^2}}}{\left( {\frac{{\partial \phi }}{{\partial \varphi }}} \right)^2} \hfill \\
  \left. { + 2\left( {{a_{12}}\left| {{\Delta _1}} \right|\left| {{\Delta _2}} \right| + {c_{11}}{{\left| {{\Delta _1}} \right|}^3}\left| {{\Delta _2}} \right| + {c_{22}}\left| {{\Delta _1}} \right|{{\left| {{\Delta _2}} \right|}^3}} \right)\cos \phi  + {c_{12}}{{\left| {{\Delta _1}} \right|}^2}{{\left| {{\Delta _2}} \right|}^2}\cos 2\phi } \right], \hfill \\ 
\end{gathered}
\end{equation}
that after the substitution of Eq. (\ref{theta_homo}) and ${\frac{{\partial \phi }}{{\partial \varphi }}}=0$ leads to Eq. (\ref{Gibbs_final}).

Minimization of the functional Eq. (\ref{Gibbs_final}) yields equations for the order parameter moduli and the phase difference $\phi$
\begin{equation}
\label{phi_homo}
\begin{gathered}
  \frac{{{k_{12}}{\hbar ^2}}}{{{R^2}}}\left| {{\Delta _1}} \right|\left| {{\Delta _2}} \right| q^2\sin \phi + 2\left( {{a_{12}}\left| {{\Delta _1}} \right|\left| {{\Delta _2}} \right| + {c_{11}}{{\left| {{\Delta _1}} \right|}^3}\left| {{\Delta _2}} \right| + {c_{22}}\left| {{\Delta _1}} \right|{{\left| {{\Delta _2}} \right|}^3}} \right)\sin \phi + 2{c_{12}}{\left| {{\Delta _1}} \right|^2}{\left| {{\Delta _2}} \right|^2}\sin 2\phi  = 0,
\end{gathered} 
\end{equation}
\begin{equation}
\label{OP1}
\begin{gathered}
  \left( {{a_{11}} + \frac{{{k_{11}}{\hbar ^2}{q^2}}}{2}} \right)\left| {{\Delta _1}} \right| + {b_{11}}{\left| {{\Delta _1}} \right|^3} + {b_{12}}\left| {{\Delta _1}} \right|{\left| {{\Delta _2}} \right|^2} + \left( {{a_{12}} + \frac{{{k_{12}}{\hbar ^2}{q^2}}}{2} + 3{c_{11}}{{\left| {{\Delta _1}} \right|}^2} + {c_{22}}{{\left| {{\Delta _2}} \right|}^2}} \right)\left| {{\Delta _2}} \right|\cos \phi  \hfill \\
   + {c_{12}}\left| {{\Delta _1}} \right|{\left| {{\Delta _2}} \right|^2}\cos 2\phi  = 0, \hfill \\ 
\end{gathered}
\end{equation}
\begin{equation}
\label{OP2}
\begin{gathered}
  \left( {{a_{22}} + \frac{{{k_{22}}{\hbar ^2}{q^2}}}{2}} \right)\left| {{\Delta _2}} \right| + {b_{22}}{\left| {{\Delta _2}} \right|^3} + {b_{12}}{\left| {{\Delta _1}} \right|^2}\left| {{\Delta _2}} \right| + \left( {{a_{12}} + \frac{{{k_{12}}{\hbar ^2}{q^2}}}{2} + {c_{11}}{{\left| {{\Delta _1}} \right|}^2} + 3{c_{22}}{{\left| {{\Delta _2}} \right|}^2}} \right)\left| {{\Delta _1}} \right|\cos \phi  \hfill \\
   + {c_{12}}{\left| {{\Delta _1}} \right|^2}\left| {{\Delta _2}} \right|\cos 2\phi  = 0. \hfill \\ 
\end{gathered}
\end{equation}

The structure of the linear terms in Eqs. (\ref{OP1}) and (\ref{OP2}) indicates a formal redefinition of the coefficients and their periodic dependence on the magnetic field due to the chosen topology. 

\section{Microscopic description of the critical temperature as a function of impurities}
\label{sec:C}

The expression for the critical temperature as a function of the impurity scattering rate $\Gamma$ can be obtained within the linearized Usadel equations supplemented by the self-consistent equations for the energy gaps. The procedure of the derivation for a multi-component superconductor has been described already in details in Ref.\ \onlinecite{Gurevich1}. Here we only give the final expression without showing the suppression of the critical temperature $T_c$ in respect to the critical temperature $T_{c0}$ of a clean two-band superconductor without impurities when $\Gamma=0$
\begin{equation}
\label{T_c_Usadel}
U\left( {\frac{\Gamma }{{\pi T_c}}} \right) =  - \frac{{2\left( {w\lambda \ln t + \lambda \left( {{\lambda _{11}} + {\lambda _{22}}} \right) - 2w} \right)\ln t}}{{2w\lambda \ln t + \lambda \left( {{\lambda _{11}} + {\lambda _{22}} - {\lambda _{12}} - {\lambda _{21}}} \right) - 2w}},
\end{equation}
where we have introduced the new function $U\left( x \right) = \psi \left( {\frac{1}{2} + x} \right) - \psi \left( {\frac{1}{2}} \right)$ expressed via  the digamma function $\psi(x)$, $t=T_c/T_{c0}$, $\lambda$ is the largest eigenvalue of the matrix of intra- and interband coefficients and $w = \det {\lambda _{ij}} = {\lambda _{11}}{\lambda _{22}} - {\lambda _{12}}{\lambda _{21}}$.
\begin{figure}
\includegraphics[width=0.70\columnwidth]{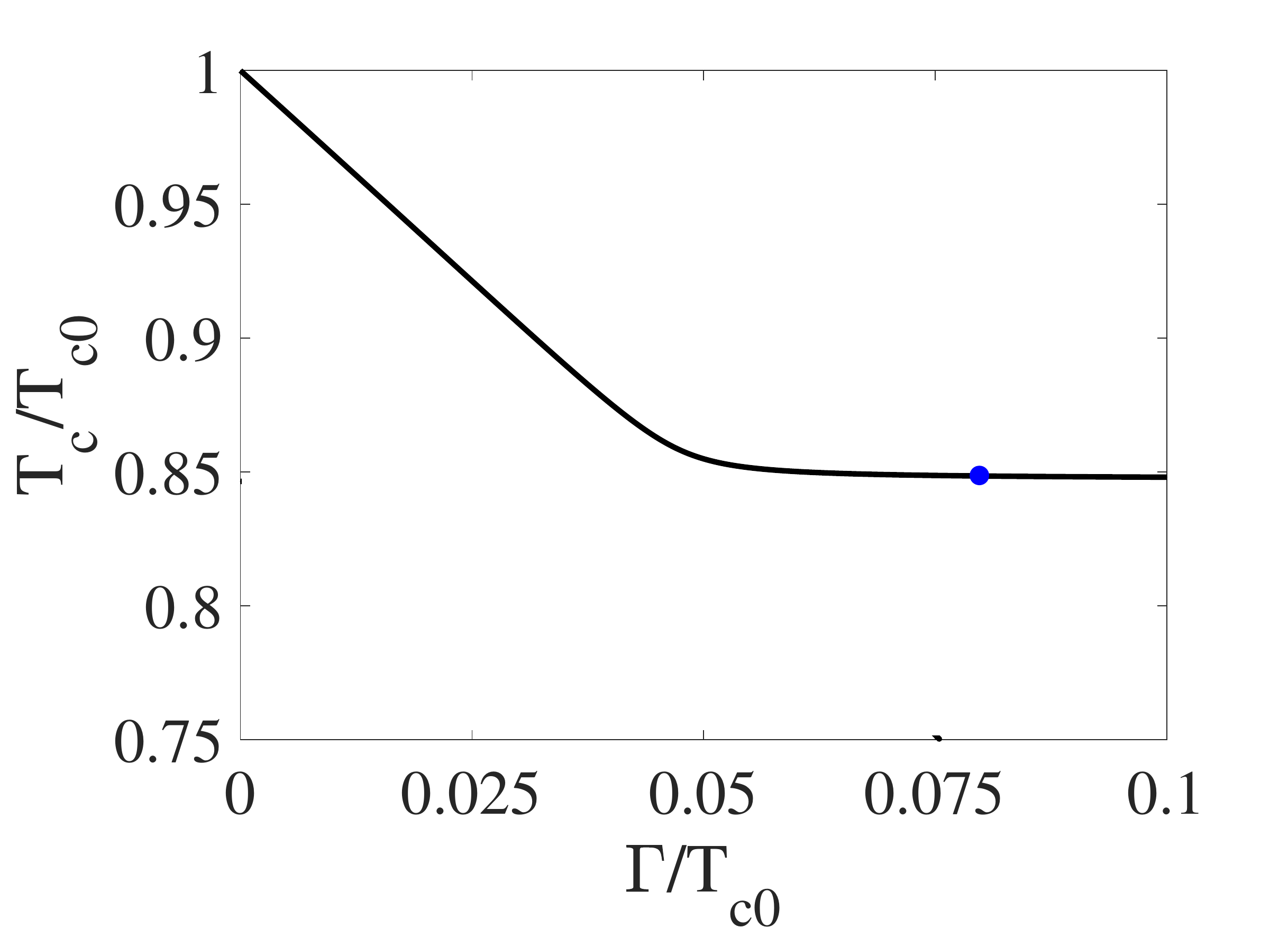}
\caption {The critical temperature $T_c$ of a dirty two-band superconductor as a function of the interband scattering rate $\Gamma$ with $\lambda_{11}=0.35$, $\lambda_{22}=0.347$,  $\lambda_{12}=\lambda_{21}=-0.01$. The values of $T_c$ and $\Gamma$ are calibrated to the critical temperature of a two-band superconductor without impurities $T_{c0}$ and $\Gamma = 0$, respectively. The blue dot corresponds to the value of $\Gamma=0.07982T_{c0}$ (and consequently $T_c=0.8485T_{c0}$), which is used in the main paper for the illustration of the order parameter symmetry oscillations.}
\label{Tc_vs_Gamma}
\end{figure}

The numerical solution of Eq.\ (\ref{T_c_Usadel}) is shown in Figure \ref{Tc_vs_Gamma}. For the sake of clarity, we have marked with a blue filled dot the point corresponding to the selected values
 of $\Gamma$ and $T_c$ used in the main text of the paper. 

\section{Estimate for a transition to an $s_{++}$ state in the bulk}
\label{sec:D}

Within the weak coupling approximation the critical temperature of a clean two-band superconductor is governed by rhe exponential factor containing the involved the four coupling constants $\lambda_{ij}$
see for instance Eq.\ (12) in Ref.\ \onlinecite{Efremov2011})
\begin{equation}
\label{estimation_Tc}
T_c \propto \exp(-1/\lambda_0)  \ , 
\end{equation}
where $\lambda_0=\frac{\lambda_{11}+\lambda_{22}}{2}+
\sqrt{\frac{\left(\lambda_{11}-\lambda_{22}\right)^2}{4}+\lambda_{12}\lambda_{21}}$. Assuming a constant bosonic prefactor in Eq. \ref {estimation_Tc} as well as a tiny residual interband attraction $\varepsilon \rightarrow +0$, i.e.\
$\lambda_{12}=\tilde{\lambda}_{12}+\varepsilon$ and $\lambda_{21}=\tilde{\lambda}_{21}+\varepsilon$, 
the ratio of the 
transtion temperature for a limiting $s_{++}$ state we look for is given explicitly by
\begin{equation}
\label{estimation_Tc_new}
\frac{T^{++}_c}{T^{\pm}_{c0}}\approx \exp \left(
 \frac{\lambda_{11}/2-\lambda_{22}/2-\sqrt{\frac{\left(\lambda_{11}-\lambda_{22}\right)^2}{4}+\lambda_{12}\lambda_{21}}}
 {\lambda_{11}\left(\lambda_{11}/2+\lambda_{22}/2+\sqrt{\frac{\left(\lambda_{11}-\lambda_{22}\right)^2}{4}+\lambda_{12}\lambda_{21}}\right)}\right) 
\end{equation}
thereby $\lambda_{11}>\lambda_{22}$ has been assumed for the sake of certainty in accord with the adopted parameter set in the main text. Without the auxiliary residual interband coupling we would arrive formally at a single band superconductor given by the system "1" decoupled from/coexisting with  a system "2" remaining in the normal state at $T=T^{++}_c$ . In this sense Eq. \ref{estimation_Tc_new} provides a lower bound for $T^{++}_c$ with always present residual attractive interband couplings.

\end{widetext}

\end{document}